\documentstyle[aps,prl,multicol,epsfig]{revtex}
\renewcommand{\v}[1]{{\bf #1}}

\newcommand{\s}{{\sigma}}

\newcommand{\rb}{{\bar{\rho}}}

\newcommand{\gr}{{\nabla}}
\def\eqa{\begin{eqnarray}}
\def\eea{\end{eqnarray}}
\newcommand{\eq}{\begin{equation}}
\newcommand{\ee}{\end{equation}}
\newcommand{\nn}{\nonumber\\}
\newcommand{\Eq}[1]{Eq.~(\ref{#1})}
\newcommand{\p}{\partial}

\newcommand{\ra}{\rightarrow}

\newcommand{\cL}{ {\cal L} }

\begin{document}
\draft
\widetext
\title{Superconductivity in a doped Mott insulator}
\author{Dung-Hai Lee}

\address{Department of Physics,University of California
at Berkeley, Berkeley, CA 94720, USA \\}

\maketitle

\widetext

\begin{abstract}

\rightskip 54.8pt

Starting from the d-wave RVB mean-field theory of Kotliar and Liu, we present a new, long-wavelength/low-energy
exact, treatment of gauge fluctuations. The result is a theory of gapless fermion quasiparticles coupled to
superconducting phase fluctuations. We will discuss the physical implications, and the similarity/differences with
a theory of BCS pairing with phase fluctuations.

\end{abstract}
\begin{multicols}{2}

\narrowtext

The high-temperature superconductors are doped Mott insulators. Shortly after the experimental report of high
$T_c$, Anderson proposed the ``resonating-valence-bond''(RVB) idea.\cite{anderson} According to this idea the root
of high temperature superconductivity is an insulating spin liquid of singlet pairs. Due to Coulomb blockade these
pairs are unable to move. Doping creates electronic vacancies that mobilize these pairs. Once mobile, the
valence-bond pairs can Bose condense into a superconducting state.

From the beginning the Neel-ordered state found in the
undoped cuprates presents a difficulty to the RVB idea.
Motivated by the experimental fact that the Neel order
is destroyed once the sample becomes metallic/superconductive,
there is a general hope that sufficient doping
stablizes the long-sort (doped) spin liquid.

The notion of RVB is made very attractive due to the experimental data on the ``underdoped cuprates''. For example
there exists a significant temperature range ($T^*>T>T_c$) where a pseudogap exists in the spin excitation
spectrum when there is no superconductivity. Angle-resolved photoemission spectroscopy (ARPES)\cite{shen} revealed
the fact that the size and the momentum dependence of the pseudogap is roughly the same as the d-wave gap in the
superconducting state. In addition, ARPES studies of undoped cuprates find a peak above the Mott-Hubbard gap with
a dispersion similar to that of spinons\cite{bob} in the RVB mean-field theories.\cite{affleck,kotliar}

On the theory side, implementation of RVB beyond mean-field theories\cite{kotliar,suzu} is hindered by the
necessity to treat strong gauge fluctuations.\cite{gauge,su2,wl} In the absence of a reliable treatment of the
gauge field, it is impossible to tell which features of the mean-field theory actually correspond to reality. In
this paper we revisit the Kotliar-Liu d-wave RVB mean-field theory.\cite{kotliar} What we were able to find is an
exact treatment of gauge fluctuations at long wavelength and low energy. Eqs.(\ref{main},\ref{ren}) plus the
discussions after it 1-6) are the main results of this paper.

Our starting point is the following boson-fermion representation of the t-J model:\cite{wl} \eq Z=\int
D[U]D[a_0]D[\psi^+,\psi]D[b^+,b]e^{-\int dt \cL}, \ee where \eqa
\cL&&=\sum_{ij}[\frac{J}{2}Tr(U^+_{ij}U_{ij})-ia_{0i}\delta_{ij}]
+\frac{1}{2}\sum_{ij}\psi^+_{i\s}[(\p_0+ia_{0i}\tau_z)\delta_{ij}\nn&&+JU_{ij}]\psi_{j\s}
+\sum_{ij}b^+_i[(\p_0+ia_{0i}-\mu)\delta_{ij}-t\chi_{ij}]b_j. \eea Here $i,j$ runs through the sites of a square
lattice, $\psi_{i\s}=\pmatrix{f_{i\s}\cr\epsilon_{\s\s'}f^+_{i\s'}}$, $\tau_z=\pmatrix{1&0\cr 0&-1}$,
$U_{ij}\equiv\pmatrix{-\chi_{ij}^* & \Delta_{ij}^*\cr \Delta_{ij}&\chi_{ij}}$, and the rest of the notations are
standard. In the literature $b_i$ and $f_{i\s}$ are often referred to as holon and spinon annihilation operators
respectively.

First, let us briefly review the results of mean-field theory.\cite{kotliar,suzu} For $x>0$, the mean-field
solutions are characterized by the following parameters: $ia_{0i}(t)\rightarrow \mu_f$, $b_i(t)\rightarrow b$,
$\chi_{ij}(t)\rightarrow \chi_0$, and $\Delta_{ij}(t)\rightarrow\Delta_0\eta_{ij}$. Here $\eta_{ij}=+1$ if $i,j$
are are $\hat{x}$-nearest-neighbor, and $=-1$ if $i,j$ are $\hat{y}$-nearest-neighbor. When the temperature is not
too high, both $\mu_f$ and $\chi_0$ are nonzero. Depending on the values of $b$ and $\Delta_0$ the low temperature
$x-T$ plane is divided into four regions: i) d-wave superconducting phase where $\Delta_0\ne 0$, and $b\ne 0$, ii)
the spin gap phase where $\Delta_0\ne 0$, and $b= 0$, iii) the Fermi liquid phase where $\Delta_0=0$, and $b\ne
0$, and iv) the strange metal phase where $\Delta_0=b=0$. The boundaries between these four phases are $T_{BE}(x)$
and $T_{pair}(x)$ at which $b$ and $\Delta_0$ develop expectation values respectively. In mean-field theory both
$T_{BE}(x)$ and $T_{pair}(x)$ mark phase transitions. This statement will be made invalid by fluctuations. A
schematic phase diagram can be found in, e.g., Ref.\cite{lee}.

According to the phase diagram\cite{lee} $T_{pair}(x)>T_{BE}(x)$ in the underdoped regime. Throughout this regime
there exists a temperature range $T_{BE}(x)<T<T_{pair}(x)$ in which a spin gap exists without superconductivity.
In the rest of the paper, we shall concentrate on the d-wave superconducting phase and the spin gap regime
mentioned above.

At temperatures much lower than $T_{pair}$ the important fluctuations include: i) the gapless spinon excitations
and ii) \eqa &&\chi_{ij}\ra \chi_0e^{ia_{ij}},~~ \Delta_{ij}\ra \Delta_0\eta_{ij}e^{i\theta_{sp,ij}}\nn
&&b_i\ra\sqrt{\rho_{b,i}}e^{i\theta_{b,i}}, ~~ a_{0i}\ra -i\mu_f+a_{0i}.\label{fluc} \eea Among these fluctuations
$a_{0i}$ and $a_{ij}$ act to enforce the no-double-occupancy constraint. More explicitly $a_{0i}$ enforces
$f^+_{i\s}f_{i\s}+b^+_ib_i=1$, while $a_{ij}$ ensures no net flow of the boson and fermion current. In the
following we present a new approach that treats the $a_{0i}$ and $a_{ij}$ fluctuations (hence the constraint)
exactly at long wavelength and low energy.

First let us write down a continuum action (its microscopic cutoff length is the holon separation) to capture all
the fluctuations discussed above: \eqa \cL&&=\psi_b^+[(\p_0+ia_0-iA_0)+\frac{1}{2m_b}|\v p+\v a-\v A|^2]\psi_b\nn
&&+ \frac{u_b}{2}(\psi^+_b\psi_b-\bar{\rho}_b)^2+\frac{K_{sp}}{2}|\phi^*_{sp}(\frac{\gr}{i}+2\v
a)\phi_{sp}|^2\nn&&
+\frac{1}{2u_{sp}}(\phi^*_{sp}\frac{\p_0}{i}\phi_{sp}+2a_0)^2+i\bar{\rho}_{sp}(\phi^*_{sp}\frac{\p_{0}}{i}\phi_{sp}+2a_{0})
\nn&&+iJ_{f\mu}(\phi^*_{sp}\frac{\p_{\mu}}{i}\phi_{sp}+2a_{\mu}) -i\bar{\rho}a_0+\cL_f[\Delta_0] \label{cont} \eea
In the above $\psi_b=\sqrt{\rho_b}e^{i\theta_b}\equiv \sqrt{\rho_b}\phi_b$ is the holon field,
$\phi_{sp}=e^{i\theta_{sp}}$ is the phase factor associated with the spinon pairs, $\bar{\rho}_b$ and
$\bar{\rho}_{sp}$ are the average densities of holon and spinon pairs, $\bar{\rho}$ is the density at
half-filling, $m_{b}$ is the holon effective mass, $A_{\mu}$ is the physical gauge field, $\cL_f$ is the Dirac
quasiparticle action which depends on the pairing amplitude $\Delta_0$,
$J_f=(\frac{1}{2}\sum_n\Psi^+_{n\s}\tau_z\Psi_{n\s} ,\frac{i}{2}v_F\Psi^+_{1\s}\Psi_{1\s},
\frac{i}{2}v_F\Psi^+_{2\s}\Psi_{2\s})$ is the 3-current of the spinon quasiparticles, where $\Psi_{n\s}$ ($n=1,2$)
are the spinon Nambu spinors associated with the two Dirac points respectively. We note that the spinon and
spinon-pair parts of \Eq{cont} are the action of a phase fluctuating BCS superconductor.\cite{bfn}

By substituting $\psi_b=\sqrt{\rho_b}\phi_b$
into \Eq{cont} we obtain
\eqa
\cL&&=\frac{\rho_b}{2m_b}|\phi^*_b(\frac{\gr}{i}+\v a-\v
A)\phi_b|^2+\frac{|\gr\rho_b|^2}{8m_b\rho_b}+
\frac{u_b}{2}(\rho_b-\bar{\rho}_b)^2\nn
&&+\frac{K_{sp}}{2}|\phi^*_{sp}(\frac{\gr}{i}+2\v a)\phi_{sp}|^2
+\frac{1}{2u_{sp}}(\phi^*_{sp}\frac{\p_0}{i}\phi_{sp}+2a_0)^2
\nn
&&+i\delta\rho_{b}(\phi^*_b\frac{\p_0}{i}\phi_b+a_0-A_0)
+iJ_{f\mu}(\phi^*_{sp}\frac{\p_{\mu}}{i}\phi_{sp}+2a_{\mu})\nn&&-2i\bar{\rho}_{sp}
(\phi^*_b\frac{\p_0}{i}\phi_b-A_0)
+i\bar{\rho}_{sp}\phi^*_{sp}\frac{\p_0}{i}\phi_{sp}\nn&&
-i\rb A_0+\cL_f[\Delta_0],
\eea
where $\delta\rho_{b}\equiv\rho_{b}-\bar{\rho}_{b}$. In
obtaining the above we have used the fact that due to the lattice
effect (i.e. when viewed by the vortices, density $\rb$ of bosons can be gauged
away because it corresponds to one flux quantum per plaquette)
$i\bar{\rho}_b\phi^*_{b}\frac{\p_0}{i}\phi_{b}=
-2i\bar{\rho}_{sp}\phi^*_{b}\frac{\p_0}{i}\phi_{b}$.

Next we Hubbard-Stratonavich decouple the first, the fourth, and the fifth terms to obtain \eqa
\cL&&=\frac{m_b}{2\rho_b}|\v
j_b|^2+\frac{u_b}{2}\delta\rho_b^2+\frac{|\gr\rho_b|^2}{8m_b\rho_b}+\frac{1}{2K_{sp}}|\v
j_{sp}|^2+\frac{u_{sp}}{2}\delta\rho_{sp}^2\nn &&+i(J_{sp\mu}+J_{f\mu})
(\phi^*_{sp}\frac{\p_{\mu}}{i}\phi_{sp}+2a_{\mu})\nn&&+iJ_{b\mu}
(\phi^*_b\frac{\p_{\mu}}{i}\phi_b+a_{\mu}-A_{\mu})-2i\bar{\rho}_{sp} (\phi^*_b\frac{\p_0}{i}\phi_b-A_0)\nn
&&+i\bar{\rho}_{sp}\phi^*_{sp}\frac{\p_0}{i} \phi_{sp}-i\rb A_0+\cL_f[\Delta_0]. \label{act1} \eea In \Eq{act1}
$J_{b\mu}\equiv (\delta\rho_b,\v j_b)$, $J_{sp\mu}\equiv (\delta\rho_{sp},\v j_{sp})$ where $\v j_b$, $\v j_{sp}$
and $\delta\rho_{sp}$ are the auxiliary field introduced by the Hubbard-Stratonavich transformation. Physically
$J_{b\mu}$ and $J_{sp\mu}$ are the three-currents of the holon and spinon pairs. In the following, for distances
greater than the holon separation ($\sim 1/\sqrt{x}$), we shall linearize \Eq{act1} by replacing $\rho_b$ in the
first term by $\rb_{b}$ and drop the third term. The resulting action read \eqa \cL&&=\frac{1}{2K_b}|\v
j_b|^2+\frac{u_b}{2}\delta\rho_b^2+\frac{1}{2K_{sp}}|\v j_{sp}|^2+\frac{u_{sp}}{2}\delta\rho_{sp}^2\nn&&
+i(J_{sp\mu}+J_{f\mu}) (\phi^*_{sp}\frac{\p_{\mu}}{i}\phi_{sp}+2a_{\mu})\nn&&+iJ_{b\mu}
(\phi^*_b\frac{\p_{\mu}}{i}\phi_b+a_{\mu}-A_{\mu}) +\cL_f[\Delta_0]\nn&& -2i\bar{\rho}_{sp}
(\phi^*_b\frac{\p_0}{i}\phi_b-A_0)+i\bar{\rho}_{sp}\phi^*_{sp}\frac{\p_0}{i} \phi_{sp}-i\rb A_0. \label{act11}
\eea In the above $K_b\equiv\frac{\bar{\rho}_b}{m_b}$ and $K_{sp}\equiv\frac{\bar{\rho}_{sp}}{m_{sp}}$.

Now we can integrate out $a_{\mu}$ exactly. The result is the constraint \eqa J_{b\mu}+2(J_{sp\mu}+J_{f\mu})=0,
\label{const} \eea i.e. the total three current with respect to $a_{\mu}$ is zero! Physically this is due to the
no-double-occupancy constraint which requires each lattice site to be occupied by either a holon or a spinon.
Since they both carry the same charge with respect to $a_{\mu}$, the total charge is constant at any time. As the
result the total three-current vanishes. If we define
$
J_{p\mu}\equiv (\delta\rho_p,\v j_p)\equiv \frac{1}{2} J_{b\mu},
$
\Eq{const} implies \eq J_{sp\mu}=-J_{p\mu}-J_{f\mu}. \label{sol} \ee Substitute \Eq{sol} into \Eq{act1} we obtain
\eqa \cL&&=\frac{1}{2}(\frac{4}{K_b}+\frac{1}{K_{sp}})|\v
j_p|^2+\frac{1}{2}(4u_b+u_{sp})\delta\rho_p^2\nn&&+iJ_{p\mu}
(\phi^*_p\frac{\p_{\mu}}{i}\phi_p-2A_{\mu})+\frac{1}{K_{sp}}\v j_p\cdot\v j_f+u_{sp}\delta\rho_p\rho_f\nn&&
+\frac{1}{2K_{sp}}|\v j_f|^2+\frac{u_{sp}}{2}\rho_f^2-i\bar{\rho}_{sp} (\phi^*_p\frac{\p_0}{i}\phi_p-2A_0)\nn
&&+\cL_f[\Delta_0]-i\rb A_0. \label{act2} \eea In the above \eq \phi_p\equiv \phi^*_{sp}\phi_b^2. \ee Now we can
integrate out $J_{p\mu}$ to obtain \eqa \cL&&=\frac{K_p}{2}|\phi^*_p\frac{\v\gr}{i}\phi_p-2\v
A|^2+\frac{1}{2u_p}(\phi^*_p\frac{\p_0}{i}\phi_p-2A_0)^2\nn&& -iz_j\v j_f\cdot(\phi^*_p\frac{\gr}{i}\phi_p-2\v A)
-iz_{\rho}\rho_f(\phi^*_p\frac{\p_0}{i}\phi_p-2A_0)\nn&&-i\bar{\rho}_{sp}
(\phi^*_p\frac{\p_0}{i}\phi_p-2A_0)+\cL_f'[\Delta_0]. \label{main} \eea In the above \eqa
&&K_p\equiv\frac{1}{K_{sp}^{-1}+4K_b^{-1}},~~~ u_p\equiv u_{sp}+4u_b\nn
&&z_j\equiv\frac{K_{sp}^{-1}}{K_{sp}^{-1}+4K_b^{-1}},~~~ z_{\rho}\equiv\frac{u_{sp}}{u_{sp}+4u_b}, \label{ren}
\eea and \eq \cL_f'[\Delta_0]\equiv \cL_f[\Delta_0]+\frac{2}{K_{sp}+4K_b}|\v
j_f|^2+\frac{2}{u_{sp}^{-1}+4u_b^{-1}}\rho_f^2, \label{qp1} \ee Eqs(\ref{main}-\ref{qp1}) is the main result of
this paper. Aside from the renormalization factor $z_{\rho}$ and $z_j$ and the quasiparticle interaction in
\Eq{qp1} the form of \Eq{main} agrees with that written down in Ref.\cite{bfn} based on weak-coupling
considerations. In the following we comment on several salient features/consequences of
Eqs.(\ref{main}-\ref{qp1}).

{\bf{1.Quasiparticle interaction:}} The last two terms in \Eq{qp1}
correspond to quasiparticle interaction.
Since both of them are local and have scaling
dimension 4 with respect to the free Dirac theory, they are
irrelevant in the renormalization group sense. In addition both coupling
constants $\frac{2}{K_{sp}+4K_b}$ and $\frac{2}{u_{sp}^{-1}+4u_b^{-1}}$
stays finite as $x\rightarrow 0$. (As $x\rightarrow 0$
we expect $K_b\sim x$,  while $K_{sp}$, $u_b$ and $u_{sp}$ all stay finite.)
For these reasons we believe that one can safely neglect the quasiparticle interaction.

In the superconducting phase, one can further integrate out the gaussian phase fluctuations in $\phi_{sp}$. The
generated spinon quasiparticle interaction is again local, and irrelevant. This result suggests that in the
superconducting state the spin degrees of freedom are asymptotically described by a free Dirac theory.

{\bf{2.Quasiparticle charge and current renormali- zation:}} An important effect of the no-double-occupancy
constraint is to introduce renormalization factors $z_{\rho}$ and $z_j$. We emphasize that when $z_j\ne 1$ and
$z_{\rho}\ne 1$, \Eq{main} is not equivalent to a BCS superconductor with phase fluctuations.\cite{bfn,note} Given
the $x$-dependence of $K_{b,sp}$ and $u_{b,sp}$, it is easy to show that $z_j\sim x$ while $z_{\rho}\sim 1$. With
such $x$-dependent $z_j$ \Eq{main} does not have the desired form written down in Ref.\cite{wl2} to explain the
temperature derivative of superfluid density. Indeed, according to the present result the suppression of the
superfluid density due to thermal quasiparticle current fluctuation will be proportional to
$x^2T$.\cite{millis,lee} It is interesting to note that since $z_{\rho}\sim 1$ we expect an x-independent linear
temperature correction to $u_p$ due to quasiparticle screening.

It is worth emphasizing that in
deriving \Eq{main} we have not assumed holon condensation, and
that Eqs.(\ref{main},\ref{ren},\ref{qp1}) are applicable in both
$\phi_p$-coherent and $\phi_p$-incoherent phases.

{\bf{3.Flux quantization:}} After integrating out
$a_{\mu}$, only the combination $\phi_p=\phi_b^2\phi^*_{sp}$ of the holon
and the spinon-pair phase factors appear in \Eq{main}. We emphasize
that the appearance of $\phi_b^2$ in the above expression
is not due to holon pairing.  This is important because
in the presence of Coulomb interaction, and in the absence
of screening at distances smaller than inter-holon separation, such
pairing will be highly energetically
unfavorable. The reason that $\phi_b^2$ appears in $\phi_p$ is
the fact that the movement of a spinon pair always causes
two holons to relocate. The combination $(\phi^*_p\frac{\p_{\mu}}{i}\phi_p-2A_{\mu})$
in \Eq{main} (note the coefficient $2$ in front of $A_{\mu}$)
implies that in the superconducting state (i.e. $\phi_p$
orders) magnetic flux will be quantized in units of $hc/2e$.

{\bf{4.The pseudogap phase:}} Under the present framework the superconducting state corresponds to the
$\phi_p$-ordered phase, but how to view the pseudogap regime? According to the RVB mean-field theory, such phase
arises from holon uncondensing. This remains qualitatively true in the present results. For small $x$ where the
superfluid density is low, there is a wide temperature range $T_c<T<T_{pair}$ in which $\phi_p$ is thermally
disordered while the spinons remain paired. In this temperature range there is no superconductivity but shows a
d-wave gap for spin excitations. Clearly, it is tempting to associate this crossover regime with the pseudogap
regime seen experimentally.

Due to
the wide separation between $T_c$ ($\sim x$) and $T_{pair}$, there
could be intermediate temperatures at which the thermal
correlation length of $\phi_p$ is comparable with the
microscopic cutoff $\sim 1/\sqrt{x}$ while is still much larger
than the spinon pairing length $\sim v_F/\Delta_0$. At these temperatures the
``bare'' superfluid density will vanish without destroying the
spin pseudogap.\cite{joe}

In the literature there exists a debate as to whether the pseudogap regime can be described as a phase-disordered
superconductor.\cite{seb,steve,y,lee} \Eq{main} gives an affirmative answer to this question. However, as pointed
out earlier, due to $z_j,z_{\rho}\ne 1$ and the discussion below, \Eq{main} does not describe an ordinary BCS
superconductor with phase fluctuations.

Lastly, since we have integrated out the gauge field exactly, the issue of whether the gauge fluctuations
destabilize the pseudogap regime\cite{uben} does not arise.\cite{kw}

{\bf{5.Spin-charge separation:}} \Eq{main} exhibits spin-charge separation in the sense that the low-energy spinon
excitations are largely determined by the RVB mean-field theory, and are little affected by the gauge
fluctuations. On the other hand the charge response depends on quantity such as $K_p$ which is greatly affected by
the gauge fluctuations. A concrete example of this separation is that the vortex core size will be very different
from spinon pairing length scale ($\sim v_F/\Delta_0$).

To calculate the
size of vortex core we restore the $\frac{|\gr\rho_b|^2}{8m_b\rho_b}$
term in \Eq{act1}. Repeating the calculations between \Eq{act11} to
\Eq{main} we obtain the following (most relevant) addition
to \Eq{main}:
\eq
\Delta\cL=\frac{1}{2m_b\bar{\rho}_b}|\gr\rho_p|^2,
\ee
Using this result we obtain
a vortex core size given by \eq
\frac{1}{\xi^2}=\frac{u_{sp}+4u_b}{\frac{1}{m_b\bar{\rho}_b}
}\sim x.
\ee
Thus as $x\ra 0$ the vortex core size diverge as
$\frac{1}{\sqrt{x}}$. In sharp contrast, the size of spinon pair ($v_F/\Delta_0$)
does not depend on $x$. A further manifestation of the spin-charge separation is in the
structure of vortex core described in the following.

{\bf{6.Vortex core:}} Near a vortex core $\rb_b$ become spatial dependent so that $\rb_b(\v x)\ra 0$ toward the
center of the vortex. Under such condition $K_b$ in \Eq{ren} becomes spatial dependent and $K_b(\v
x)=\frac{\rb_b(\v x)}{m_b}\ra 0$ toward the center of the vortex. As the result both $K_p$ and $z_j$ vanishes in
the vortex core. On the contrary the spinon excitations (which depend on the mean-field pairing amplitude
$\Delta_0$) remain largely unaffected by the depletion of superfluid density.\cite{ng} (To be more precise, since
holons are depleted from the vortex core, the RVB mean-field pairing amplitude $\Delta_0$ should slightly increase
toward the center of the vortex.) This is qualitatively different from the behavior found in BCS superconductors
where the pairing amplitude (or local gap) collapse at the center of the vortex. This difference could be
manifested in the absence of vortex-core induced midgap states.\cite{expt}

At closing it is important to point out that the treatment in this paper eventually becomes invalid as $x\ra 0$,
where a charge SU(2) symmetry emerges.\cite{su2,wl} We believe that the enlarged gauge symmetry is responsible for
antiferromagnetism. Moreover, the SU(2) gauge fluctuations at length scales smaller than the holon separation can
cause important modifications of the bare parameters (or even introducing new spinon interaction terms) in our
theory. Finally, the cutoff length and energy scales in this work can be affected by the (dynamic) stripe
correlation\cite{kiv} presents in the underdoped systems.
\\

\noindent{\bf{Acknowledgement}} I thank G. Baskaran and Steve Kivelson for useful discussions, and Xiao-Gang Wen
for helpful comments. This work was supported by NSF grant DMR 99-71503.
\\

\centerline{\bf BIBLIOGRAPHY}

\bibliographystyle{unsrt}

\begin{thebibliography}{99}
\bibitem{note} By performing the gauge transformation
leading from ``nodon'' to electron in Ref.\cite{bfn}, it is
impossible to trade the coupling between fermion and the $\phi_p$
in \Eq{main} with an order parameter phase.
\end{thebibliography}

\end{multicols}
\end{document}